\documentclass[11pt]{article}

\usepackage{graphicx}
\usepackage{hyperref}
\usepackage{amsfonts}
\usepackage{cite}
\usepackage{comment}

\begin{document}

\begin{center}

{\bf\LARGE Comparison of QCD Curves \\ [5mm] with Elastic  $pp$ Scattering Data }
\\
\vspace{4cm}  

H.M. FRIED
\\
{\em Department of Physics \\
Brown University \\
Providence R.I. 02912 USA}\\
fried@het.brown.edu\\
[5mm]
P.H. TSANG
\\
{\em Department of Physics \\
Brown University \\
Providence R.I. 02912 USA}\\
peter\_tsang@brown.edu\\
[5mm]
Y. GABELLINI
\\
{\em Institut de Physique de Nice\\
UMR 7010 CNRS\\
Site Sophia
06560 Valbonne France}\\
yves.gabellini@inphyni.cnrs.fr\\
[5mm]
T. GRANDOU
\\
{\em Institut de Physique de Nice\\
UMR 7010 CNRS\\
Site Sophia
06560 Valbonne France}\\
Thierry.Grandou@inphyni.cnrs.fr\\
[5mm]
Y.M. SHEU
\\
{\em Department of Physics \\
Brown University \\
Providence R.I. 02912 USA}\\
ymsheu@alumni.brown.edu

\newpage

\vskip1truecm {\Large Abstract}
\end{center}

\indent Using previously described functional techniques for some non--perturbative, gauge invariant, renormalized QCD processes, a simplified version of the amplitudes --- in which forms akin to Pomerons naturally appear --- provides fits to ISR and LHC--TOTEM $pp$  elastic scattering data. Those amplitudes rely on a specific function $\varphi (\vec b)$ which describes the fluctuations of the transverse position of quarks inside hadrons.

{\section{Introduction}}

In a set of recent papers \cite{qcd1,qcd2,qcd3,qcd4} that we will call for all along this article, the present authors have shown how it is possible to proceed from any relativistic, gauge dependent generating functional of QCD, to new explicit solutions for its quark and/or antiquark correlation functions, in a gauge invariant and non--perturbative manner. All those computations rely strongly on a given function, $\varphi (\vec b)$, partly phenomenological, which depicts the transverse fluctuations of quark position inside a hadron \cite{qcd1,qcd5}. The most salient applications of these newly discovered techniques for computing 2n--point Green functions  include the derivation of quark binding \cite{qcd2} and nucleon binding \cite{qcd3} potentials, these two derivations being summarized in a review article \cite{qcd_summary}. The present paper aims to keep on investigating the application of those methods, in order to proceed from those correlation functions describing  hadron scattering to S matrix elements for high--energy, elastic, proton--proton scattering. This procedure involves the introduction of a few parameters determined in comparison with the forty year old Intersecting Storage Rings, ISR, experimental data \cite{isr_data1,isr_data2,isr_data3,isr_data4}, where the center of mass energy was ${\sqrt s} \simeq 40$ GeV, together with the most recent LHC--TOTEM data \cite{LHC} with ${\sqrt s} \simeq 10$ TeV.
\medskip

In Section 2, we recall three main features -- among four -- of our approach to QCD fermionic 4--point functions, leading to elastic scattering processes and potentials. 
Those are: 

\medskip
1) the existence of ``gluon bundles" in place of ordinary gluon propagator,

\medskip
2) the ``effective locality" exact property, that simplifies a lot all the computations,

\medskip
3) the introduction of the transverse fluctuation function $\varphi (\vec b)$ that we must somewhat modelize, and check that it fits the various data.
\medskip

These will be qualitatively discussed, as the quantitative elements are in the references cited 
above \cite{qcd1,qcd2,qcd3,qcd4}.
\medskip

In Section 3,  we set the theoretical frame in which we will do our computation, and come back to the fourth main feature of our non--perturbative approach to QCD: the renormalization program \cite{qcd4}, and set the gluon bundle renormalization conditions specific for the calculation of this elastic $pp$ scattering differential cross section.
\medskip

In Section 4, we recall the eikonal amplitudes obtained in Refs.\cite{qcd1} and \cite{qcd2} for gluon bundle exchange and in Refs.\cite{qcd3} and \cite{qcd4} for quark loop chain exchange, these two processes being needed to obtain the $pp$ elastic cross section. We give some insight of how to go from our QCD formula to a computationally simpler expression.

In Section 5, we give our  ``postdiction" of elastic $\displaystyle\frac{d\sigma}{dt}$ to be compared to the ISR and TOTEM data. We also give the expression for the total elastic cross section.
\medskip

\begin{figure}
\centering
\includegraphics[width=12cm]{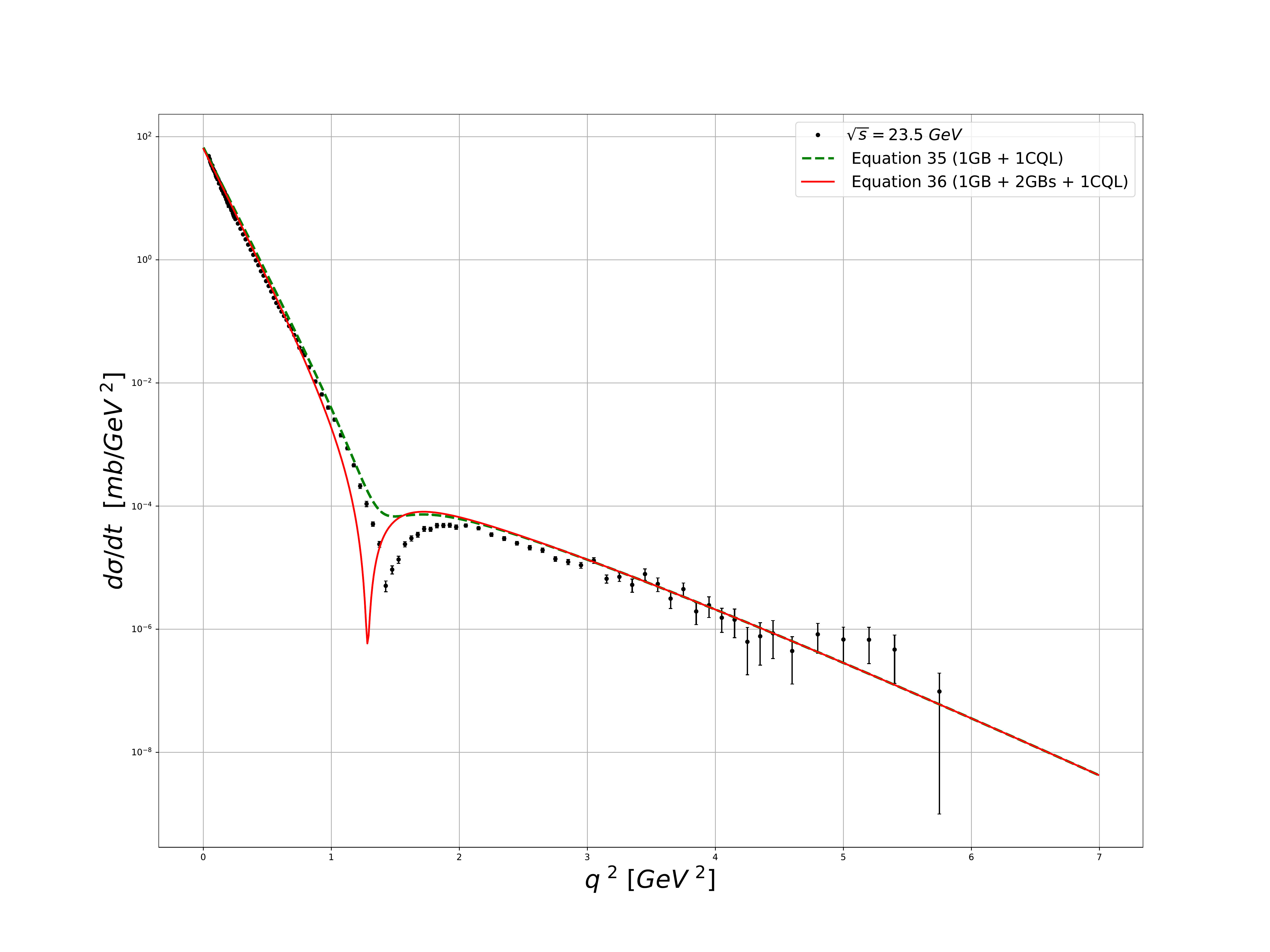}
\caption{Elastic pp scattering differential cross section at $\sqrt{s} = 23.5$ GeV. Black dots are experimental data, dashed line is the result of eq.(35), solid line comes from eq.(36). }
\label{24gev}       
\end{figure}

\begin{figure}
\centering
\includegraphics[width=12cm]{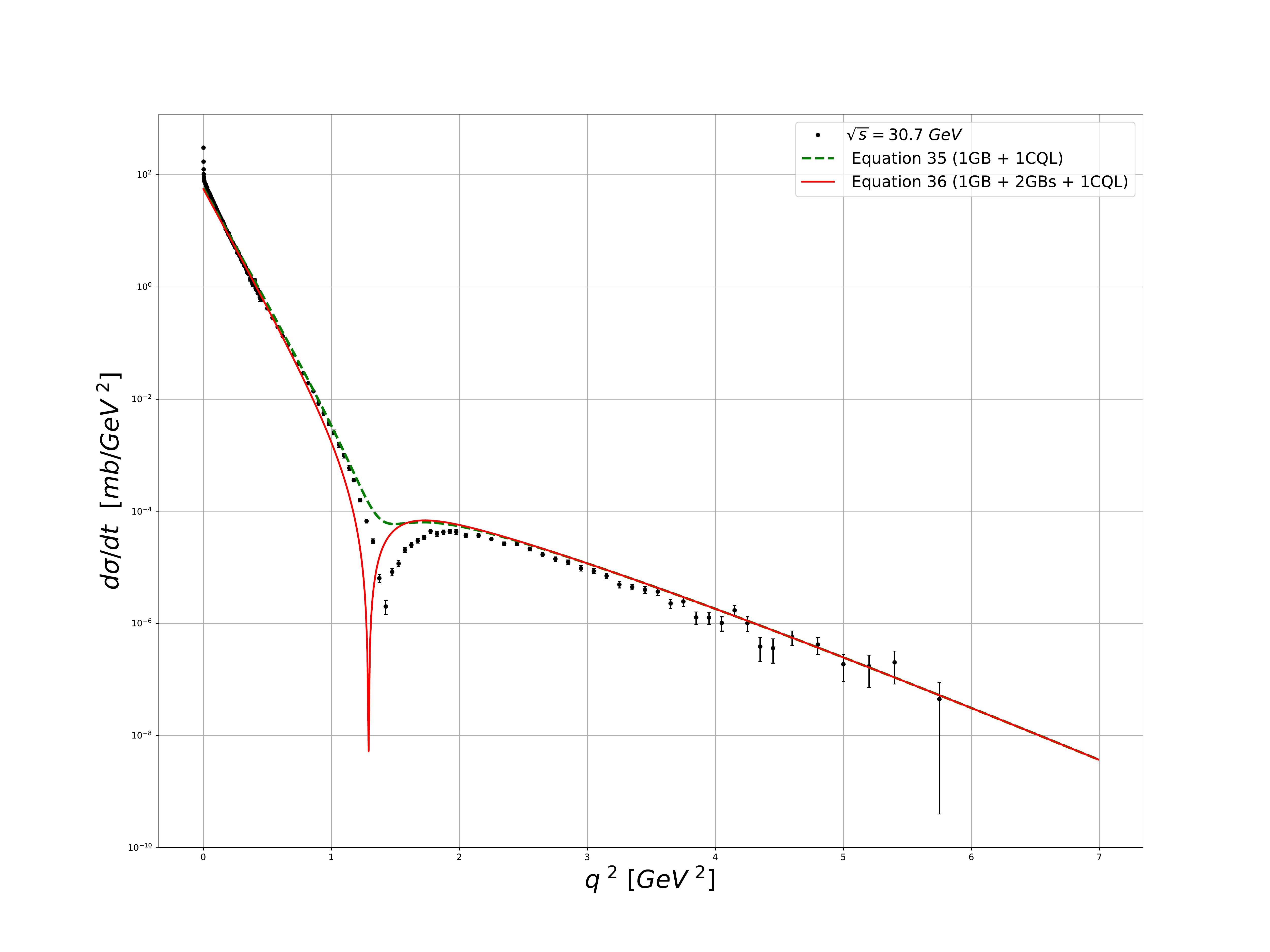}
\caption{Elastic pp scattering differential cross section at $\sqrt{s} = 30.7$ GeV. Black dots are experimental data, dashed line is the result of eq.(35), solid line comes from eq.(36). }
\label{31gev}
\end{figure}

\begin{figure}
\centering
\includegraphics[width=12cm]{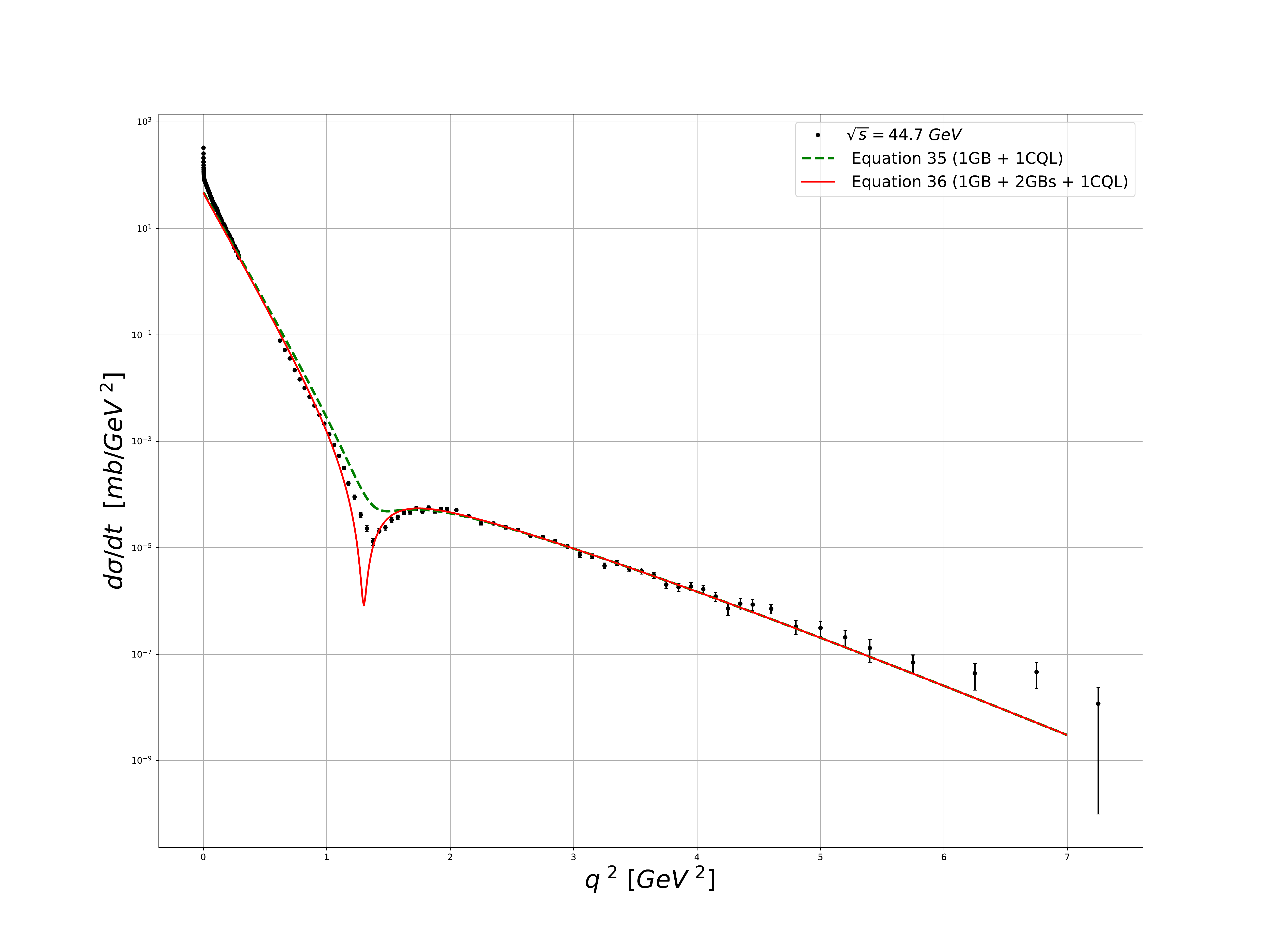}
\caption{Elastic pp scattering differential cross section at $\sqrt{s} = 44.7$ GeV. Black dots are experimental data, dashed line is the result of eq.(35), solid line comes from eq.(36). }
\label{45gev}       
\end{figure}

\begin{figure}
\centering
\includegraphics[width=12cm]{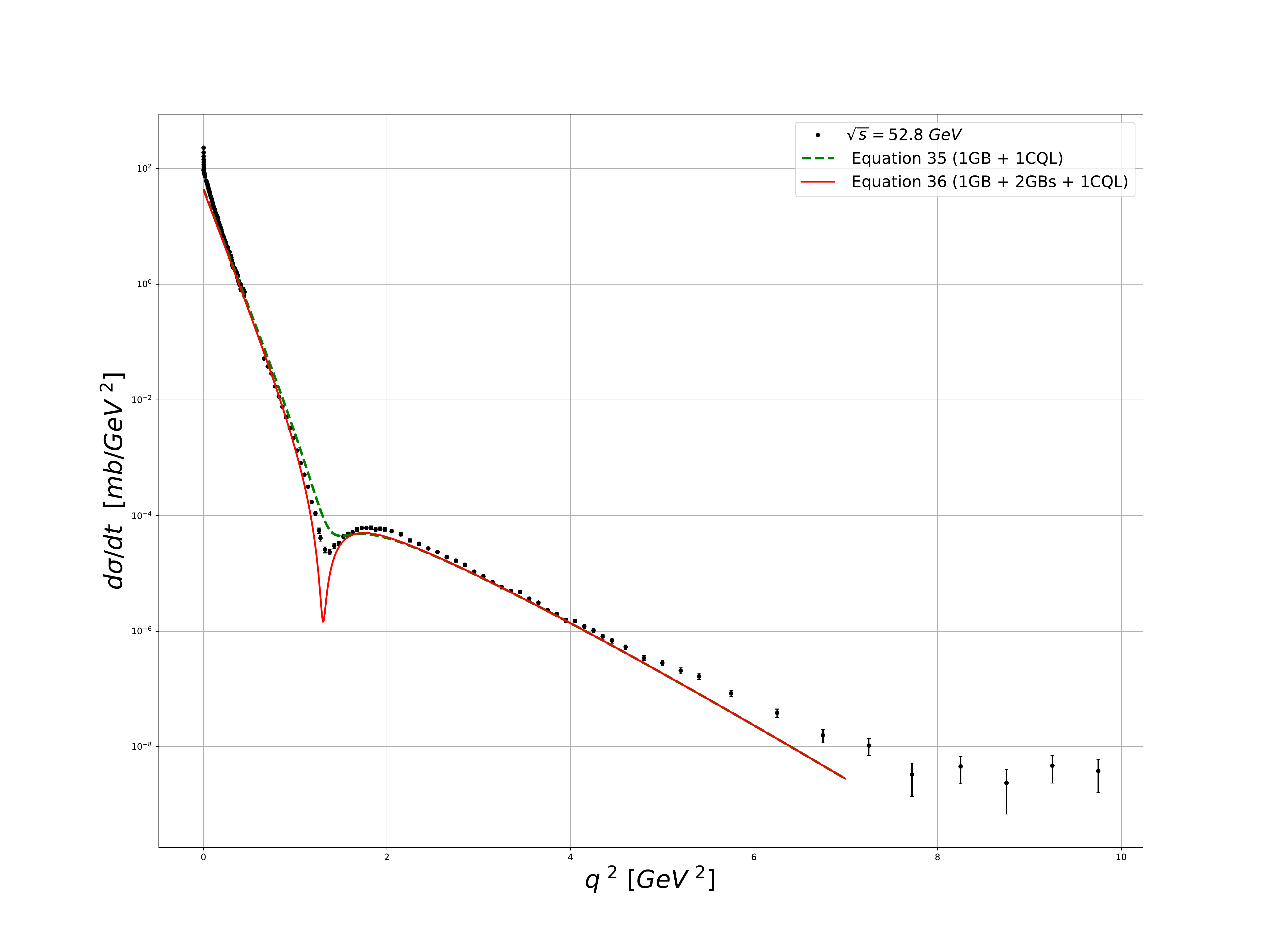}
\caption{Elastic pp scattering differential cross section at $\sqrt{s} = 52.8$ GeV. Black dots are experimental data, dashed line is the result of eq.(35), solid line comes from eq.(36). }
\label{53gev}       
\end{figure}

\begin{figure}
\centering
\includegraphics[width=12cm]{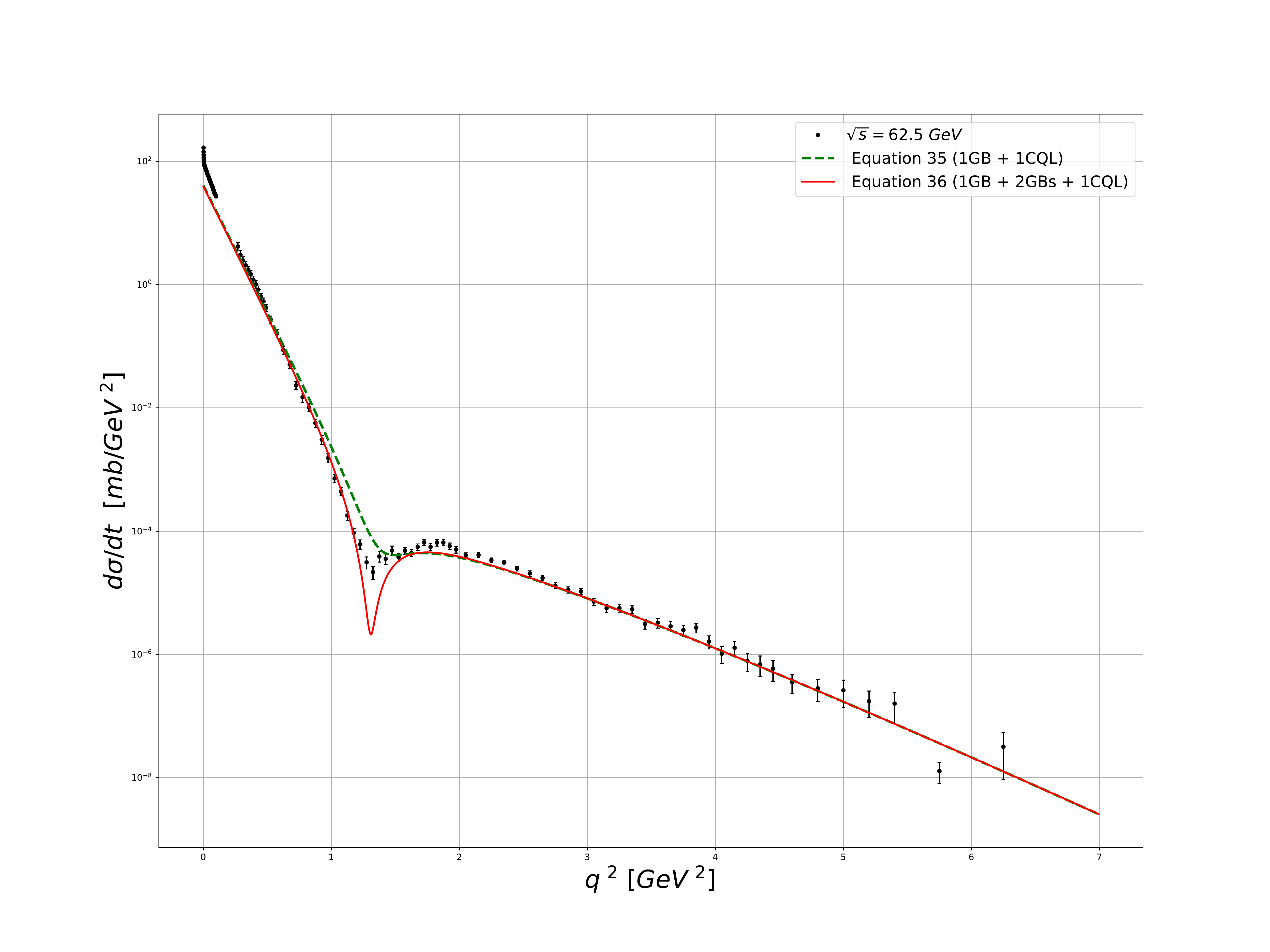}
\caption{Elastic pp scattering differential cross section at $\sqrt{s} = 62.5$ GeV. Black dots are experimental data, dashed line is the result of eq.(35), solid line comes from eq.(36). }
\label{63gev}      
\end{figure}

Section 6 is the Summary. We discuss the results of our simple model, and how we could improve our predictions for elastic cross sections, in particular at higher energies. We also make a brief reference to the Pomeron theory \cite{Regge,pomeron,diffraction}: as can be seen from the fits of Figures \ref{24gev}-\ref{63gev}, we have, in effect, proposed field theory derivations of what might be called the non--perturbative QCD Pomerons \cite{DHM,DL}.

\medskip

In Appendix A, a justification is produced on the angular average used in eq.(8). 
\medskip

In Appendix B, we give some insight about the energy dependence of our theoretical cross sections. 

\bigskip

{\section{Non Perturbative QCD and its 4--fermion amplitude}}

Let us start by mentioning the general thrust of our approach, which is to begin with the QCD generating functional $\displaystyle\mathcal{Z}_{\mathrm{QCD}}[j, \bar{\eta}, \eta]$ \cite{qcd1,qcd2,qcd3,qcd4}, with gluons in any gauge; and then perform a simple rearrangement which brings this generating functional into a completely gauge invariant form ( see  Ref.\cite{qcd1}, Appendix C ). At that stage, the exact functional operations required are made possible due to a gaussian relation written and used by M. Halpern \cite{halpern1,halpern2}. Choosing to write $\mathcal{Z}$ in the  Feynman gauge, we obtain: 

\begin{equation}
\matrix{\displaystyle\mathcal{Z}_{\mathrm{QCD}}[j, \bar{\eta}, \eta] & = \displaystyle\mathcal{N} \,e^{\displaystyle\,\frac{i}{2} \int{j\,\mathbf{D}_{\mathrm{F}}^{(0)}j} }\, \int{\mathrm{d}[\chi]\, \, e^{\displaystyle\,\frac{i}{4} \int\chi^{2} } }\, e^{\displaystyle\,\mathfrak{D}_{A}^{(0)}}  \, e^{\displaystyle\,\frac{i}{2} \int{\chi\,\mathbf{F} } }\hfill\cr\noalign{\medskip} \hfill\times & e^{\displaystyle\, \frac{i}{2} \int{ A \,\Bigl(\mathbf{D}_{\mathrm{F}}^{(0)}\Bigr)^{-1} A} } \, e^{\displaystyle\,i\int{\bar{\eta}\, \mathbf{G}_{\mathrm{F}}[A] \,\eta} + \mathbf{L}[A]}\Bigr|_{A = \int{\mathbf{D}_{\mathrm{F}}^{(0)} j} }\hfill}
\end{equation}
where $j, \bar{\eta}, \eta$ are the gluon, quark and antiquark sources, $A_\mu^a$ the eight gluon fields, $a =  1, \ldots ,8$, $\mathbf{F}_{\mu \nu}^{a} = \partial_{\mu} A_{\nu}^{a} -  \partial_{\nu} A_{\mu}^{a} + g f^{abc} A_{\mu}^{b} A_{\nu}^{c}$ the gluon strenght tensor, $f^{abc}$ being the antisymmetric $\mathrm{SU_c(3)}$ structure constants, $\mathbf{D}_{\mathrm{F}}^{(0)}$ the free gluon Feynman propagator, with $\Bigl(\mathbf{D}_{\mathrm{F}}^{(0)}\Bigr)^{-1}\Bigr|_{\mu\nu}^{ab} = -g_{\mu\nu}\,\delta^{ab}\,\partial^{2}$, $\mathbf{G}_{\mathrm{F}}[A]$ the Feynman quark Green's function: $\mathbf{G}_{F}[A] = [m + \gamma\,(\partial - i g\lambda A)]^{-1}$, $\mathbf{L}[A]$ the closed quark loop functional: $\mathbf{L}[A] = {\mathrm{Tr}} \ln{\bigl[ 1 - ig\gamma \lambda A \mathbf{S}_{\mathrm{F}} \bigr]}$, $\lambda$ being the Gell--Mann matrices, $\mathbf{S}_{\mathrm{F}} = \mathbf{G}_{\mathrm{F}}[0]$, $\exp{[\mathfrak{D}_{A}^{(0)}]}$ is the linkage operator with $\displaystyle \mathfrak{D}_{A}^{(0)} = - \frac{i}{2} \int{\frac{\delta}{\delta A}\,\mathbf{D}_{\mathrm{F}}^{(0)}\frac{\delta}{\delta A}}$ and, last but not least, $\chi_{\mu\nu}^a$ are the Halpern auxiliary fields, antisymmetric in their Lorentz indices.
\medskip

The process we are interested in is quark--quark and/or quark--antiquark elastic scattering. Its amplitude is given by ( Ref.\cite{qcd2}, eq.(20) ):

\begin{equation}
\matrix{\displaystyle\mathbf{M}(x_{1}, y_{1}; x_{2}, y_{2}) & = \displaystyle\frac{\delta}{\delta \bar{\eta}(y_{1})}\frac{\delta}{\delta \eta(x_{1})} \frac{\delta}{\delta \bar{\eta}(y_{2})} \frac{\delta}{\delta \eta(x_{2})}\,\mathcal{Z}[j, \bar{\eta}, \eta]\Bigr|_{\eta=\bar{\eta}=0; j=0} \hfill\cr\noalign{\medskip} \hfill & = \displaystyle\, \mathcal{N} \, \int{d[\chi] \,\, e^{\,\displaystyle{\frac{i}{4}\int{\chi^{2}}}} \, e^{\,\displaystyle{\mathfrak{D}_{A}^{(0)}}} \,}e^{\,\displaystyle{\frac{i}{2}\int{\chi\,\mathbf{F}} +\frac{i}{2}\int{A\left(\mathbf{D}_{F}^{(0)}\right)^{-1}A }}}\hfill\cr\noalign{\medskip} \hfill &\times \,\mathbf{G}_{\mathrm{F}}(x_{1}, y_{1}|gA) \, \mathbf{G}_{\mathrm{F}}(x_{2}, y_{2}|gA) \, e^{\,\displaystyle{\mathbf{L}[A]} }\big|_{A=0} \quad -\{ {1} \leftrightarrow {2} \}\hfill}
\end{equation}

The result of the functional operation $e^{\displaystyle\,\mathfrak{D}_{A}^{(0)}}$ is then the appearance of, first, a new QCD quantity that we call ``gluon bundle", second, a new and exact property called ``effective locality"   and third, the explicit demonstration of gauge independence for this process. 

These results ensue from the following formula, demonstrated, for instance, in \cite{HMF}:

\begin{equation}
\matrix{&\displaystyle e^{\displaystyle\,-{i\over2}\!\int\!\!d^4x\,d^4y\,{\delta\over\delta A_{\mu}^a(x)}D_{\mu\nu}^{ab}(x,y){\delta\over\delta A_{\nu}^b(y)}}\,e^{\displaystyle\, {i\over2}\!\int\!\!d^4x\,d^4y\,A_{\mu}^a(x)K_{\mu\nu}^{ab}(x,y)A_{\nu}^b(y)\, +\,{i}\!\int\!\!d^4x\,Q_{\mu}^a(x)A_{\mu}^b(x)}\Bigr|_{A =0 }\cr\noalign{\smallskip} &= \displaystyle e^{\displaystyle\, {i\over2}\!\int\!\!d^4x\,d^4y\,Q_{\mu}^a(x)\bigl[D(1-KD)^{-1}\bigr]_{\mu\nu}^{ab}(x,y)\,Q_{\nu}^b(y)}\,e^{\displaystyle\, -{1\over2}{\rm Tr}\ln(1-KD)}\hfill}
\end{equation}

that leads, using the ingredients of relation (2) in both quenched -- $\mathbf{L}[A] = 0$ -- and eikonal limits ( see eq.(25) to eq.(29) in Ref.\cite{qcd1} ) to:

\begin{equation}
\bigl[D(1-KD)^{-1}\bigr]_{\mu\nu}^{ab}(x,y) = - (gf\!\cdot\!\chi)^{-1}\Bigr|_{\mu\nu}^{ab}(x)\,\delta^{(4)}(x-y)
\end{equation}

where $(f\!\cdot\!\chi)_{\mu\nu}^{ab} = f^{abc}\chi^c_{\mu\nu}$.
\medskip

There, in eq.(4), the major features of our computation mentioned above appear:
\medskip

1) the ``gluon bundle" manifests itself in the $(gf\!\cdot\!\chi)^{-1}(x)$ function, in place of a $D(x-y)$ gluon propagator.
\medskip

2) the ``effective locality" shows up: the kernel $\bigl[D(1-KD)^{-1}\bigr](x,y)$ is proportional to $\delta^{(4)}(x-y)$. 

\noindent And its main consequences are:

\hskip1cm i) the replacement of the functional integral $\int{\mathrm{d}}[\chi]$ in eq.(2) by a set of ordinary Lebesgue integrals, which can be evaluated exactly but are easily estimated using pencil and paper. 

\hskip1cm ii) the necessity to introduce a smooth `` transverse fluctuation function '' $\varphi (\vec b)$ ( Ref.\cite{qcd1}, eq.(44) and paragraph above )  in place of a singular Dirac distribution $\delta^{(2)}(\vec b)$, $\vec b$ being the impact parameter between the two scattering quarks in the center of mass frame that the effective locality property forces to be equal to zero. See Refs.\cite{qcd5,EL0,EL1,EL2} for rigourous results on effective locality and $\varphi (\vec b)$.
\medskip

3) The left hand side of eq.(4) depends on a choice of gauge. The right hand side doesn't.

\medskip
Instead of working in the quenched and eikonal limits, we could have used the  Fradkin representation for  $\mathbf{G}_{\mathrm{F}}[A]$ and $\mathbf{L}[A]$ \cite{fradkin1,fradkin2}, without any approximations, and obtain the same three features as above ( see section 2 of Ref.\cite{qcd2} for instance ).
\medskip

As a conclusion for this section and as derived in our QCD papers, we have found that some radiative corrections to the correlation functions can be obtained by the exchange of ``gluon bundles"  between any pair of quarks and/or antiquarks, including quarks which form virtual, closed quark loops, and those which are, or are about to be bound into hadrons. Each gluon bundle consists of a sum over an infinite number of virtual gluons, with space--time and color indices properly maintained and displayed. We were then able to define quark binding potentials ( without the use of static quarks )( see Ref.\cite{qcd2} ), and to produce a qualitative nucleon binding potential, in which two nucleons form a model deuteron ( see Ref.\cite{qcd3} ).

The calculations can in principle all be defined and carried through exactly, and in a finite manner; but for simplicity and ease of presentation, certain obvious approximations were presented. These simplifications are retained in the present paper, in which the above analysis is applied to the ISR elastic scattering of two protons, at a variety of energies in the 10 GeV range \cite{peterthesis} and to the TOTEM elastic scattering in the 10 TeV range. There will appear below an additional set of simple approximations to specific integrals, again for reasons of subsequent simplicity. 

\bigskip

{\section{Elastic $pp$ Scattering and Gluon Bundle Renormalization}}

\begin{figure}
\centering
\includegraphics[width=12cm]{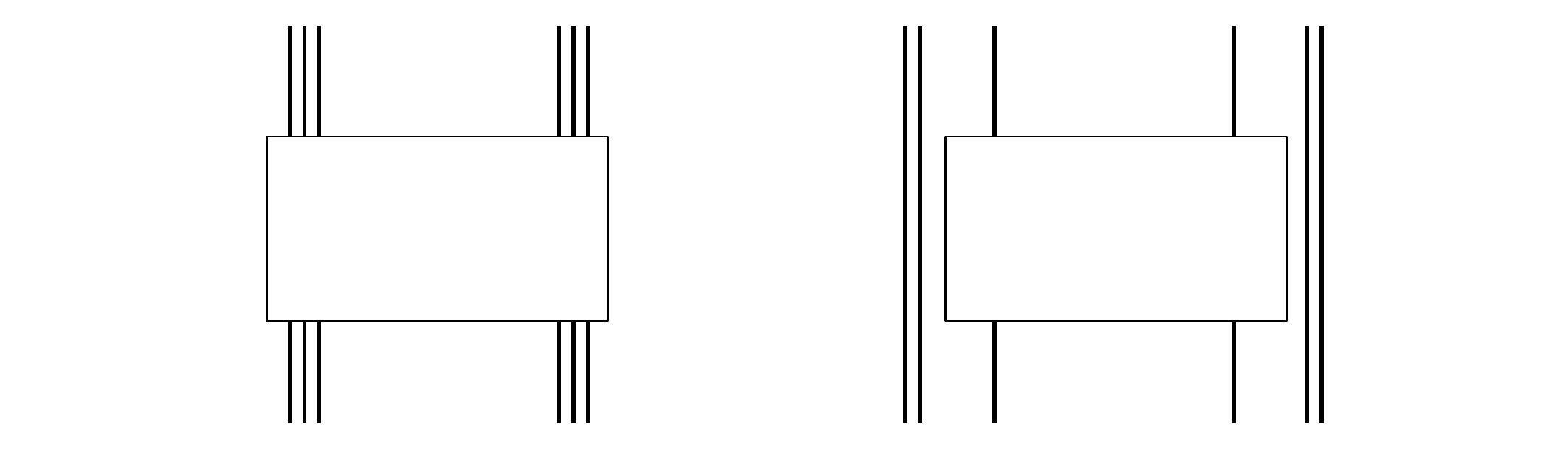}
\caption{The elastic pp scattering. On the left, the six-body interaction; on the right, the two-body approximation.}
\label{ppscat}       
\end{figure}

	We emphasize that these descriptions of elastic $pp$ scattering can, in principle, be evaluated exactly in terms of six--body quark interactions ( we leave aside the gluon and sea content of the protons ), using Random Matrix methods \cite{qcd5,EL1,EL2,random_matrices}, but in order to keep this paper one of finite length, we have employed several approximations when evaluating relevant integrals. Perhaps the most serious simplification has been performed at the very beginning, by assuming that the scattering is "truly elastic", so that each triad of scattering valence quarks remain bound into its initial proton during the entire scattering process. This precludes, for example, the interchange of any quarks comprising each proton, as well as other more complicated possibilities, and can be clearly incorrect as energies increase. But it does replace a six--body quark problem by a two--body scattering problem (Fig.\ref{ppscat}); and the corrections to this two--body approximation are easily and intuitively defined, by the insertion of a weak energy dependence, phenomenologically obtained from the data. While it is important to understand that the correlation functions of our QCD functional procedure can be exactly calculated, it is surely a computational and physical advantage to employ the two--body approximation, which seems to work rather well at ISR and LHC energies.
	 	
\begin{figure}
\centering
\includegraphics[width=12cm]{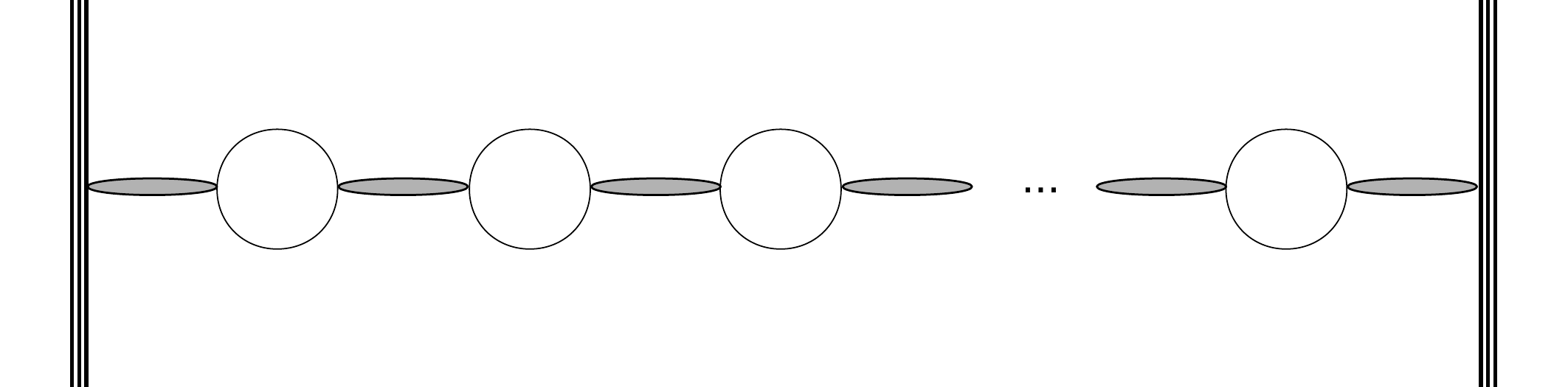}
\caption{Renormalization adopted such that no more than two gluon-bundles are attached to each quark loop; and each chain ends on a quark bound into a hadron.}
\label{chainloop}      
\end{figure}

Concerning renormalization for quark and gluon bundle interactions, there is no hint, no previous problem to which one can turn for intuitive assistance; rather the question of gluon bundle renormalization may, in part, be decided by subsequent simplicity, and with the parameters of that renormalization fixed by the data. That passage from correlation functions to S matrix elements was described in paper \cite{qcd4}, in which non--perturbative quark and gluon bundle renormalization was defined. In this formulation, one doesn't consider processes with individual gluons, and conventional perturbative renormalization is here redefined in terms of gluon bundles interacting with quark loops, and with quarks forming hadronic bound states. A special and surely the simplest form of renormalization was adopted, in which quark loops automatically appear only in chains, with no more than two gluon bundles attached to each loop; and each chain ends on a quark bound into a hadron, as in Fig.\ref{chainloop}.

	We now turn to a detailed treatment of gluon bundle renormalization, specific to the present problem, in which each proton is a bound state of three quarks, with these three quarks here interacting with the three quarks of the other proton. It should be understood that even at low energies we are completely neglecting electrodynamic effects, and quark spin effects, which can always be added separately.
	From the original definition of the Halpern functional integral, plus the appearance of effective locality, at each end of a gluon bundle there appears a quantity $\delta$, which divides into two classes: those which connect to a quark which is, or is about to be bound into a hadron; and those which connect to a quark loop. Before renormalization, each of these $\delta$ must vanish; but renormalization here means that: 
\medskip

	1) for quarks of the first group, the "physical particle" of QCD, the $\delta$ is a finite quantity $\delta_q(E)$.  Each $\delta_q(E)$ has a dimension, which we may think of as time, or distance; and thanks to the Heisenberg inequality, the natural choice is to replace that $\delta_q$ by a dimensionless constant multiplying $1/E$, even though this leads to a rapid decrease of the differential cross section as the energy increases. But as the energy increases to ISR and LHC values, one finds that cross section for all scatterings is about the same, although still decreasing, but very slowly. The reason is presumably the onset and continued growth of  "quasi--state" processes: there are more and more ways of interchanging quarks and combining quarks and loops to produce a final state of two protons. The shape of the $q^2$ dependence of the ISR and LHC scatterings is barely affected, and this is presumably due to the fact that however complicated the intermediate "quasi--states" might be, the end product of each elastic process must be two protons.

	Following this interpretation, we must now change to a specific form of $\delta_q(E)$, one which permits a very slow decrease with increasing energy; and for this we have chosen 

\begin{equation}
\delta_q(E) \propto (1/m)(m/E)^p = (\lambda/m)(m/E)^p
\end{equation}

 where $\lambda$ and $p$ ( $0\!<\!p\!<\!1$ ) to be chosen by the data and $m$ on order of the pion mass. Of course, this is a phenomenological choice of the variation with energy of all amplitudes so constructed in the ISR and LHC range, and seems to be the best one can do under the two-body restrictions. See Appendix B for a more thorough discussion.
\medskip

	2) the $\delta$ at the quark loop end of the gluon bundle is to vanish. Combined with the expected UV log divergence of the loop, $\ell$, this gives a finite -- and small -- dimensionless $\kappa$ parameter: 

\begin{equation}
 \delta^2 \ell = {\kappa\over \bar m^2} 
\end{equation}

$\kappa$ and $\bar m$, the mass associated to the chain, are real parameters, extracted from the data.

\bigskip

{\section{Summarizing our previous results}}

In preparation for the computation of the elastic $pp$ differential cross section, let's recall the results obtained for the exchange between two quarks of 1) gluon bundles and 2) quark loop chains.
\medskip

We start by writing the eikonal representation of the scattering amplitude:

\begin{equation}
\displaystyle T(s,\vec q) =\frac{is}{2M^{2}}\int d^{2}b\ e^{\displaystyle i\vec q\cdot \vec b}\ [1-e^{\displaystyle i{\mathbb X}(s,\vec b)}]
\end{equation}
where ${\mathbb X}(s,\vec b)$ is the eikonal function appropriate to the scattering, when $s=4{\cal E}^2$, where ${\cal E}$ is the center of mass energy of each incident proton, $\vec b$ is the impact parameter of the collision in the center of mass frame, $\vec q$ is the momentum transfer in that frame: $\vec q^{\,2} = |t |<< s$, and $M$ the mass of that proton. In this simplifying frame, we neglect the spin, the angular momentum and, as state below, the color dependence of the quarks involved in the scattering.

Let's consider the centerpiece of eq.(7) and its integral:
\begin{equation}
e^{\displaystyle i{\mathbb X}(s,\vec b)} = N \int d[\chi]\, e^{\displaystyle i/4 \int\! \chi^2}[\det(f\!\cdot\!\chi)^{-1}]^{1/2}\ \mathcal{F}\Big((f\!\cdot\!\chi)\Big)
\end{equation}
derived for instance in Ref.\cite{qcd1}, eq.(33), Ref.\cite{qcd2}, eq.(36) and Ref.\cite{qcd3}, eq.(21) from the original Halpern functional integral. The $\mathcal{F}$ we use, explicited below  in relation (13) and (17), represent the exponential of the gluon bundles (13) and of the quark loop chain (17) exchanged between each valence quark of each proton (with appropriate and hidden statements of the binding of each triad, which are to be understood). 

We recall that $((gf\!\cdot\!\chi)^{-1})^{ab}_{\mu\nu}$ is the quantity characterizing the gluon bundle and its locality property: 
\begin{equation}
<\!x|(gf\!\cdot\!\chi)^{-1}|y\!>\, = (gf\!\cdot\!\chi(x))^{-1}\,\delta^{(4)}(x-y)
\end{equation}

We now make two additional approximations in the evaluation of integral (8):

\indent Eq.(8) can be expressed in calculable form by the introduction of Random Matrix Methods ~\cite{qcd5,EL2,random_matrices}. But, we prefer, first, thanks to the eikonal kinematics, to replace the Lorentz $\mu\nu$ indices of $f\!\cdot\!\chi$ by a single pair $03$, and then, second, replace $(f\!\cdot\!\chi)^{ab}_{03}$ by $R$, where $R^2$ denotes the magnitude of $(f\!\cdot\!\chi)^2$, and all of its color--angular integrations are supressed. This last simplification assumes that the color--angular integrations over different color coordinates have no real bearing on the dynamical outcome of (8); and that the important part of the exact (8) will depend only on the magnitudes of $f\!\cdot\!\chi$. A justification of this simplifying assumption is given in Appendix A. 

\medskip

\hskip1truecm 1) The gluon bundle exchange in the eikonal limit.

\medskip
Let's outline the results obtained in Ref.\cite{qcd1} and \cite{qcd2}, in the eikonal and $\mathbf{L}[A] \equiv 0$ limits.

\medskip
That is, we rewrite (8) in the form:

\begin{equation}
e^{\displaystyle i{\mathbb X}(s,\vec b)} =  N'\int_0^{\infty} \!R^7 dR\, e^{\displaystyle (i/4) R^2}\,R^{-4}\,\mathcal{F}(R)
\end{equation}
where the measure $\prod_a d[\chi^a]$ has been replaced by its radial part and the determinant factor of (8) has been replaced by $R^{-4}$,  and $N'$ is the new normalization constant such that:

\begin{equation}
  N'\int_0^{\infty} \!dR \,R^3\ e^{\displaystyle(i/4) R^2}=1
  \end{equation}
  
Integral (11) can be performed and yields: $N'=-1/8$.

Before giving $\mathcal{F}(R)$, let's come back to $\varphi (\vec b)$ mentioned in section 2, which, in fact, is the most important ingredient of our calculation.
It has been established (  Ref.\cite{qcd1}, eq.(39) ) that the effective locality property appearing in the quark--quark and/or quark--antiquark elastic scattering produces in the exponential a term proportional to $\delta^{(2)}(\vec b)$, where we remind that $\vec b$ is the impact parameter of the collision in the center of mass frame. Of course, such a singular term gives no contribution to the amplitude. So, we choose to replace it by a normalized gaussian  function (  Refs.\cite{qcd3} and \cite{EL0} for instance ), centered around $\vec b = 0$ with a range on the order of $1/m$, $m$ being related to the pion mass:
\begin{equation}
\varphi (\vec b) = {m^2\over\pi}\, e^{\displaystyle -m^2b^2}
\end{equation}
This choice is physically reasonable, mathematically tractable under a Fourier transform and compatible with the data.
As mentioned in the previous section, due to our choice of renormalization, there are two $\delta$ parameters in our processes. The one associated with the gluon bundle is $\delta_q$ ( see eq.(5) ).

We now give an explicit expression for eq.(10). The eikonal for gluon bundle exchanges is given, for instance, in Ref.\cite{qcd2} eq(58), where we have:

\begin{equation}
\mathcal{F}^{(G.B.)}(R) =  \,e^{\displaystyle{i g\, \delta_q^2\, \varphi(\vec b)R^{-1}}}
\end{equation}
so that:
\begin{equation}
e^{\displaystyle i{\mathbb X}^{(G.B.)}(s,\vec b)} =  N'\int_0^{\infty} \! dR\, R^3\,e^{\displaystyle (i/4) R^2}\,e^{\displaystyle{i g\, \delta_q^2\, \varphi(\vec b)R^{-1}}}
\end{equation}

We notice that the $R=0$ lower bound causes the $\mathcal{F}^{(G.B.)}(R)$ to oscillate infinitely rapidly, and thus makes no contribution to the integral.

\medskip
\hskip1truecm 2) The quark loop chain  exchange in the eikonal limit.

\medskip
Let's summarize the results obtained in Ref.\cite{qcd3} and \cite{qcd4}, where we have kept the closed quark loop functional $\mathbf{L}[A]$ but discarded the simple gluon bundle ( no loops ) exchanges. And consider the amplitude with  a single loop between two gluon bundles.
 
We start again from eq.(8):

$$e^{\displaystyle i{\mathbb X}(s,\vec b)} = N \int d[\chi]\, e^{\displaystyle i/4 \int\! \chi^2}[\det(f\!\cdot\!\chi)^{-1}]^{1/2}\ \mathcal{F}\Big((f\!\cdot\!\chi)\Big)$$
but this time, the $\mathcal{F}$ we use, explicited below  in relations (17) and (18), represents the exponential of a quark loop chain exchanged between each valence quark of each proton, see Ref.\cite{qcd3}, eq.(21). 

Due to the presence of two gluon bundles in the one loop chain, we get two $f\!\cdot\!\chi$ functions, at two different space-time points $x_1$ and $x_2$, leading to their respective magnitudes $R_1$ and $R_2$, and we obtain the eikonal function in the form:

\begin{equation}
e^{\displaystyle i{\mathbb X}(s,\vec b)} =  N''\!\!\int_0^{\infty}\!\! dR_1\,R_1^3\, e^{\displaystyle (i/4) R_1^2}\int_0^{\infty}\!\! dR_2\,R_2^3\, e^{\displaystyle (i/4) R_2^2}\,\mathcal{F}(R_1,R_2)
\end{equation}
$N''$ being the new normalization constant.
Every bundle comes with a $\varphi(\vec b)$ function, and by making the convolution product of the two of them we obtain a new $\bar\varphi(\vec b)$ ( see Ref.\cite{qcd3} eq.(26) ):

\begin{equation}
\bar\varphi (\vec b) = {\bar m^2\over2\pi}\, e^{\displaystyle -{\bar m^2\over2}b^2}
\end{equation}

The mass $\bar m$ involved in the quark loop chain exchange needs not to be the same as the one in the gluon bundle. The presence of the quark loop has for effect to introduce a laplacian in front of $\bar\varphi $, and, taking into account the two $\delta_q(E)$ ( see eq.(5) ) and $\kappa/\bar m^2$ ( see eq.(6) ), we get (  Ref.\cite{qcd3}, eq.(40) ) for quark loop chain contributions:

\begin{equation}
\mathcal{F}^{(Q.L.C.)}(R_1, R_2) = e^{\displaystyle -i {\mathcal{C}(\vec b, E)\over R_1R_2}}
\end{equation}
with :

\begin{equation}
\mathcal{C}(\vec b, E) = - g^2 \delta_q^2\, (\kappa/\bar m^2)\, \Delta \bar\varphi(\vec b)
\end{equation}

so that:

\begin{equation}
e^{\displaystyle i{\mathbb X}^{(Q.L.C.)}(s,\vec b)} =  N''\!\!\int_0^{\infty}\!\! dR_1\,R_1^3\, e^{\displaystyle (i/4) R_1^2}\int_0^{\infty}\!\! dR_2\,R_2^3\, e^{\displaystyle (i/4) R_2^2}\,e^{\displaystyle i g^2 \delta_q^2\, (\kappa/\bar m^2)\, \Delta \bar\varphi(\vec b)(R_1R_2)^{-1}}
\end{equation}

As for the bundle case, the $R_1$ and $R_2$ integrals give no divergences in the 0 limit.

\medskip
The second simplification employed for these amplitudes is in the evaluation of the integration over the $R$ magnitudes of (14) and the $R_1$ and $R_2$ magnitudes of (19). 

We use the following approximation scheme, by introducing a $\beta$ parameter, that will be set equal to $1/4$ at the end:

\begin{equation}
e^{\displaystyle i{\mathbb X}(s,\vec b)} =  N'\int_0^{\infty} \!dR\,R^3 e^{\displaystyle i\beta R^2}\mathcal{F}(R) = N'\Big(\!-i \frac{\partial}{\partial \beta}\Big)\int_0^{\infty}\!dR\,R \  e^{\displaystyle i\beta R^2}\ \mathcal{F}(R)
\end{equation}
and with the variable change:

\begin{equation}
R^2=iu\ ,\ \ \  R = \sqrt{iu}
\end{equation}
we obtain:

\begin{equation}
e^{\displaystyle i{\mathbb X}(s,\vec b)} =  \frac{N'}{2} \Big(\frac{\partial}{\partial \beta}\Big)\int_0^{\infty} du\ e^{\displaystyle -\beta u}\ \mathcal{F}\Big({\sqrt{iu}}\Big)
\end{equation}

The integral of (22) has serious contributions only for $u < 1/\beta$ which we approximate as:

\begin{equation}
e^{\displaystyle i{\mathbb X}(s,\vec b)} =  \frac{N'}{2} \Big(\frac{\partial}{\partial \beta}\Big)\int_0^{1/\beta} du\ \mathcal{F}\Big({\sqrt{iu}}\Big)= \mathcal{F}\Big(\sqrt{i/\beta}\,\Big) = \mathcal{F}\Big(R =2\sqrt{i}\,\Big)
\end{equation}

Using formla (23) we obtain:

\begin{equation}
e^{\displaystyle i{\mathbb X}^{(G.B.)}(s,\vec b)} =  e^{\displaystyle{\sqrt{i} g\, \delta_q^2\, \varphi(\vec b)/2}}
\end{equation}

Using twice the same trick for the $R_1$ and $R_2$  integrals, we obtain:

\begin{equation}
e^{\displaystyle i{\mathbb X}^{(Q.L.C.)}(s,\vec b)} =  \,e^{\displaystyle{\,g^2\, \delta_q^2\, (\kappa/\bar m^2)\, \Delta \bar\varphi(\vec b)/4}}
\end{equation}

We again emphasize that our functional representations can, in principle, all be calculated exactly; but to keep this paper more easily readable, we have resorted to these approximations of this Section. 

\bigskip

{\section{Evaluating the QCD elastic differential and total cross sections}}

1) The differential cross sections.

\medskip
We now take into account the possibility of exchanges of both gluon bundles and quark loop chains between the two scattering quarks involved in the $pp$ elastic collisions. We remind that a sketch of the differential cross section computed below has already been given in Ref.\cite{qcd4}, section 4. 

We recall the eikonal representation of the scattering amplitude (eq.(7)), with its associated differential cross section:

\begin{equation}
\displaystyle T(s,\vec q) =\frac{is}{2M^{2}}\int d^{2}b\ e^{\displaystyle i\vec q\cdot \vec b}\ [1-e^{\displaystyle i{\mathbb X}^{pp}(s,\vec b)}]
\end{equation}

\begin{equation}
\displaystyle\frac{d\sigma}{dt} = \frac{M^4}{\pi s^2}\, | T |^2
\end{equation}

where ${\mathbb X}^{pp}(s,\vec b)$ is the eikonal function appropriate to the $pp$ elastic scattering. 

We can now rely on Ref.\cite{qcd2} eq.(35) and Ref.\cite{qcd3} eq.(15) to (21) to obtain:

\begin{equation}
e^{\displaystyle i{\mathbb X}^{pp}(s,\vec b)} = e^{\displaystyle i{\mathbb X}^{(G.B.)}(s,\vec b)}\,e^{\displaystyle i{\mathbb X}^{(Q.L.C.)}(s,\vec b)}
\end{equation}
 
This leads to an amplitude of:
\begin{equation}
T(s,\vec q) =\frac{is}{2M^{2}}\int d^{2}b\ e^{\displaystyle i\vec q\cdot \vec b}\ \Big[\,1- e^{\displaystyle{\sqrt{i} g\, \delta_q^2\, \varphi(\vec b)/2}} \,e^{\displaystyle{\,g^2\, \delta_q^2\, (\kappa/\bar m^2)\, \Delta \bar\varphi(\vec b)/4}}\,\Big]
\end{equation}

giving a differential cross-section of:
\begin{equation}
\displaystyle\frac{d\sigma}{dt}  = \frac{1}{4\pi }\, \Bigg| \int d^{2}b\ e^{\displaystyle i\vec q\cdot \vec b}\ \Big[\,1- e^{\displaystyle{\sqrt{i} g\, \delta_q^2\, \varphi(\vec b)/2}} \,e^{\displaystyle{\,g^2\, \delta_q^2\, (\kappa/\bar m^2)\, \Delta \bar\varphi(\vec b)/4}}\,\Big] \Bigg|^2
\end{equation}

For the ISR and LHC amplitudes, which are decreasing with increasing energy, it is appropriate to expand those exponentials, retaining only its one and one plus two gluon bundle portion, for the first exponential, plus one loop chain, for the second, which means that will give two families of curves ( see Fig.\ref{gluonbundle} and Fig.\ref{oneloop} below ). Other terms in that expansion would produce correspondingly smaller corrections.  Of course, we need to check that both exponents are small, which will be the case. For instance: 
$${1\over2}\,g\, \delta_q^2\, \varphi(\vec b) < {1\over2}\,g\, \delta_q^2\, \varphi(0) = {g\over2\pi}\,\lambda^2\,\Big(\frac{m}{E}\Big)^{2p} < 1$$
for the ISR and LHC data. The same for the other exponent, thanks to the smallness of the $\kappa$ parameter.

\medskip

\begin{figure}
\centering
\includegraphics[width=10cm]{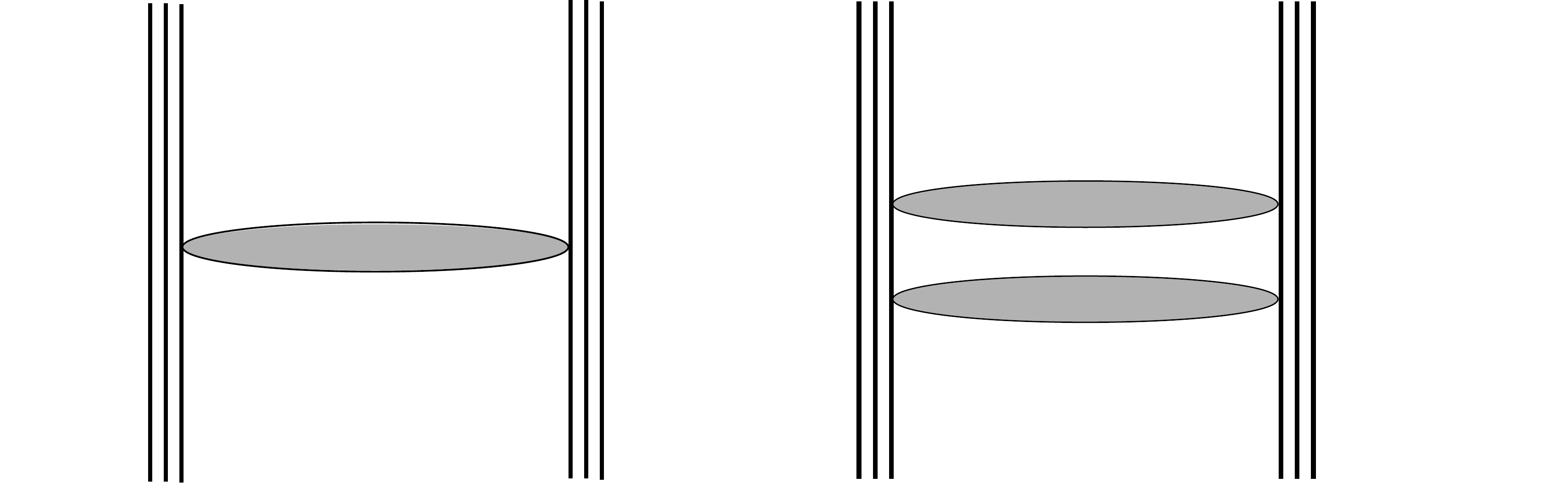}
\caption{The one and two gluon-bundle exchange as the first ingredients for our scattering amplitudes. }
\label{gluonbundle}       
\end{figure}
\begin{figure}
\centering
\includegraphics[width=6cm]{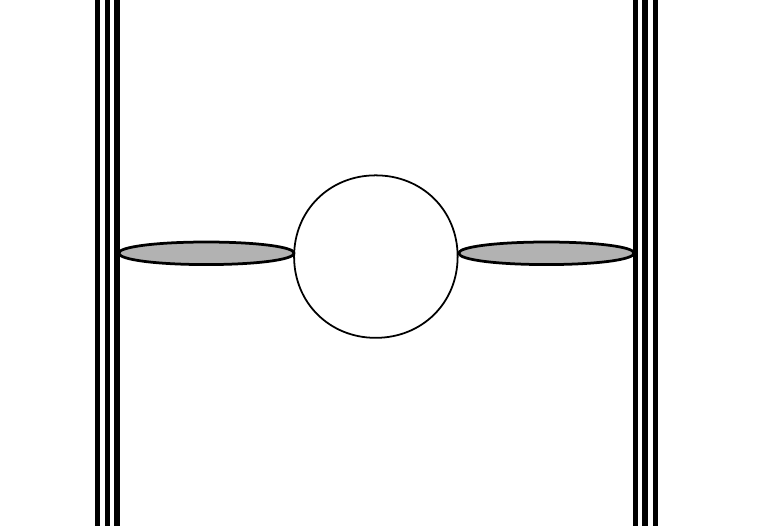}
\caption{The single quark loop chain as the second ingredient for our scattering amplitudes.}
\label{oneloop}       
\end{figure}

\medskip

Expanding eq.(29), we get for one gluon bundle plus one quark loop chain exchange:

\begin{equation}
T_1(s,\vec q) =-\frac{is}{2M^{2}}\int d^{2}b\ e^{\displaystyle i\vec q\cdot \vec b}\ \Big[\,\sqrt{i}\,{g\over2}\, \delta_q^2\, \varphi(\vec b)  + \,{g^2\over4}\, \delta_q^2\, (\kappa/\bar m^2)\, \Delta \bar\varphi(\vec b)\,\Big]
\end{equation}

and for an additional two gluon bundle exchange: 
\begin{equation}
T_2(s,\vec q) =-\frac{is}{2M^{2}}\int d^{2}b\ e^{\displaystyle i\vec q\cdot \vec b}\ \Big[\,\sqrt{i}\,{g\over2}\, \delta_q^2\, \varphi(\vec b)  + \,{1\over2}\,i\,{g^2\over4}\, \delta_q^4\,\varphi^2(\vec b) + \,{g^2\over4}\, \delta_q^2\, (\kappa/\bar m^2)\, \Delta \bar\varphi(\vec b)\,\Big]
\end{equation}

The $b$ integration is straightforward.  The amplitude for $qq$ elastic scattering can then be written, in the case of one gluon bundle and one quark loop exchange:

\begin{equation}
T_1(s,\vec q) =\frac{s}{2M^{2}}\,{g\over2}\biggl({\lambda\over m}\biggr)^2\bigg(\frac{m}{E}\bigg)^{2p}\, \Big[\,-{1\over\sqrt2}\,e^{\displaystyle{-q^2/4m^2}}  + i \,\Bigl({1\over\sqrt2}\, e^{\displaystyle{-q^2/4m^2}} + {g\over2}\,\kappa\, {q^2\over \bar m^2}\,e^{\displaystyle{-q^2/2\bar m^2}}\,\Bigr)\,\Big]
\end{equation}

and in the case of one plus two bundles:

\begin{equation}
\matrix{\displaystyle T_2(s,\vec q) &\displaystyle =\frac{s}{2M^{2}}\,{g\over2}\biggl({\lambda\over m}\biggr)^2\bigg(\frac{m}{E}\bigg)^{2p}\, \Big[\,\Bigl(-{1\over\sqrt2}\,e^{\displaystyle{-q^2/4m^2}}  + {1\over2}\,{g\over2}\,\delta_q^2\,{m^2\over2\pi}\,e^{\displaystyle{-q^2/8m^2}}\,\Bigr)\hfill \cr\noalign{\smallskip} &\displaystyle +\, i \,\Bigl({1\over\sqrt2}\, e^{\displaystyle{-q^2/4m^2}} + {g\over2}\,\kappa\, {q^2\over \bar m^2}\,e^{\displaystyle{-q^2/2\bar m^2}}\,\Bigr)\,\Big]\hfill}
\end{equation}
where we have used: $\sqrt i = e^{\displaystyle{5i\pi/4}} $

Then, taking into account the 27 multiplicity factor that represents the number of possible quark pairs, each coming in 3 colors, going from $qq$ sub--cross section to $pp$ cross section and choosing for each quark energy $E = \displaystyle   {1\over3} {\cal E} = {\sqrt s\over6}$, our approximate formulas to represent elastic $pp$ scattering at ISR and LHC energies are:

\begin{equation}
\matrix{\displaystyle  \frac{d\sigma_1}{dt}(s,q^2) &\displaystyle = K\,{27\over4\pi}\,{g^2\over4}\,\biggl({\lambda\over m}\biggr)^4\,\bigg(\frac{6m}{\sqrt s}\bigg)^{4p}\,\bigg[ \,{1\over2}\,e^{\displaystyle{-q^2/2m^2}}\hfill \cr\noalign{\smallskip} &\displaystyle +\,\Bigl({1\over\sqrt2}\, e^{\displaystyle{-q^2/4m^2}} + {g\over2}\,\kappa\, {q^2\over \bar m^2}\,e^{\displaystyle{-q^2/2\bar m^2}}\,\Bigr)^2\,\bigg]\hfill}
\end{equation}
 and
 \begin{equation}
\matrix{\displaystyle  \frac{d\sigma_2}{dt}(s,q^2) &\displaystyle = K\,{27\over4\pi}\,{g^2\over4}\,\biggl({\lambda\over m}\biggr)^4\,\bigg(\frac{6m}{\sqrt s}\bigg)^{4p}\, \bigg[\,\Bigl({1\over\sqrt2}\,e^{\displaystyle{-q^2/4m^2}}  - \,{g\over2}\,{\lambda^2\over 4\pi}\,\bigg(\frac{6m}{\sqrt s}\bigg)^{2p}\,e^{\displaystyle{-q^2/8m^2}}\,\Bigr)^2\hfill \cr\noalign{\smallskip} &\displaystyle + \,\Bigl({1\over\sqrt2}\, e^{\displaystyle{-q^2/4m^2}} + {g\over2}\,\kappa\, {q^2\over \bar m^2}\,e^{\displaystyle{-q^2/2\bar m^2}}\,\Bigr)^2\,\bigg]\hfill}
\end{equation}

where $K$ is the usual conversion factor needed to obtain the cross section in mb, when the masses and the energies are given in GeV: $K=0.44$ mb GeV$^{-2}$.     Concerning the choice of values for the mass parameters, their inverse are proportional to the size of the tranverse gluon fluctuations between different quark lines, which quantities our approximations cannot determine, and their values must here be fixed by comparison with the data. It happens that the mass values chosen to fit the ISR and TOTEM data are intuitively clear, with the ``exterior" parameter $m$,  related to gluon bundle exchanges, much closer to a pion mass than is the larger ``interior" $\bar m$, related to chain exchanges. 
\medskip

We list below the values of the parameters of eq.(35) and (36).
\medskip

For the ISR data:

$g =7.0 $ 

$p  = 0.13$

$\lambda = 0.5$

$\kappa =  - 6.8 \,10^{-4}$

$m = 0.23$ GeV $\simeq 1.5\, m_{\pi}$

$\bar m = 0.64$ GeV $\simeq 4.5\, m_{\pi}$

\bigskip

\bigskip

For the TOTEM data:

$g = 7.0$ 

$p = 0.055$

$\lambda = 0.72$

$\kappa =  -4.2\,10^{-3}$

$m = $ 0.16 GeV $\simeq m_{\pi}$

$\bar m = 0.41$ GeV $\simeq 3\, m_{\pi}$

\bigskip

\begin{figure}
\centering
\includegraphics[width=12cm]{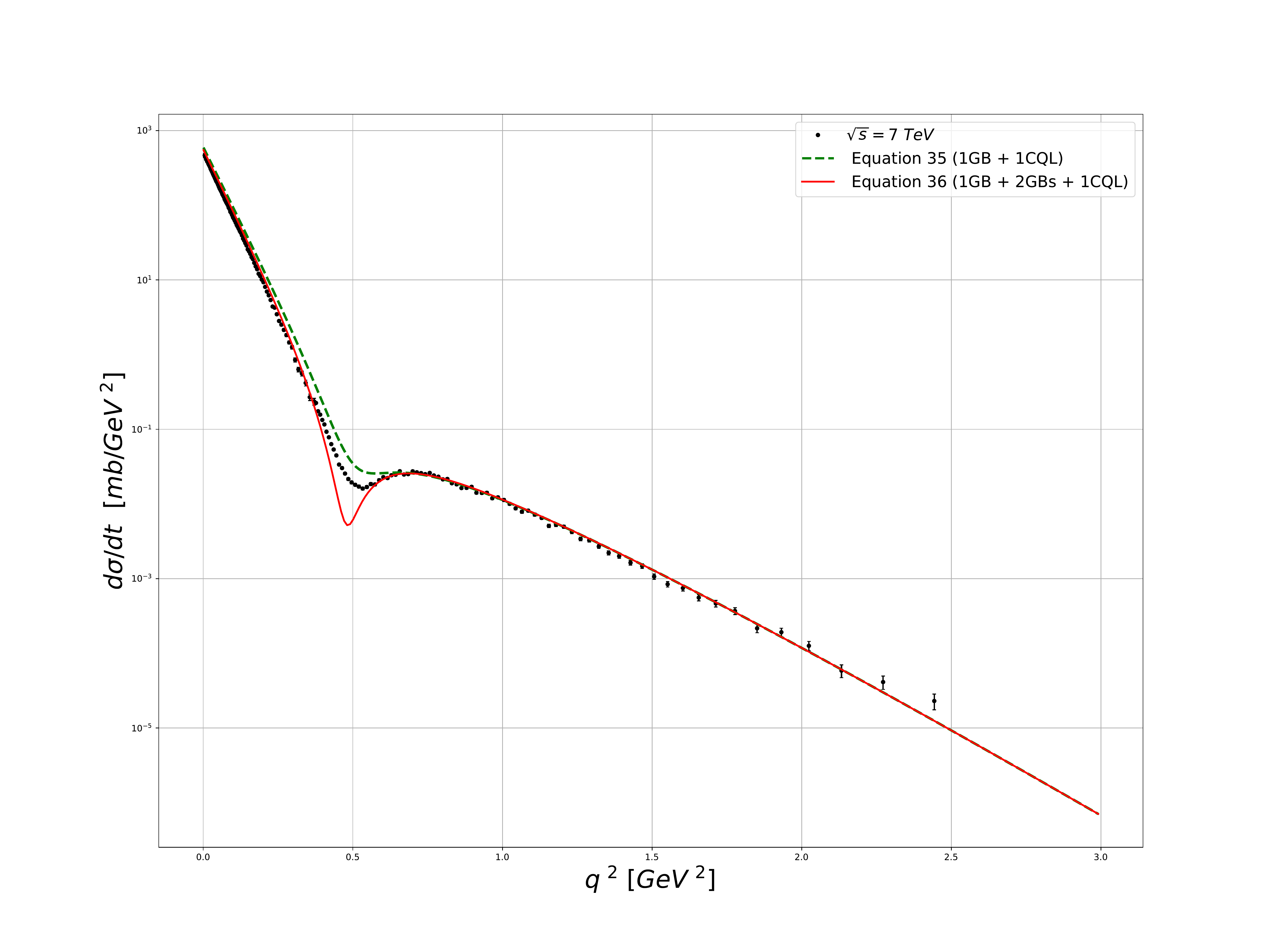}
\caption{Elastic pp scattering differential cross section at $\sqrt{s} = 7$ TeV. Black dots are experimental data, dashed line is the result of eq.(35), solid line comes from eq.(36). }
\label{7tev}       
\end{figure}

\begin{figure}
\centering
\includegraphics[width=12cm]{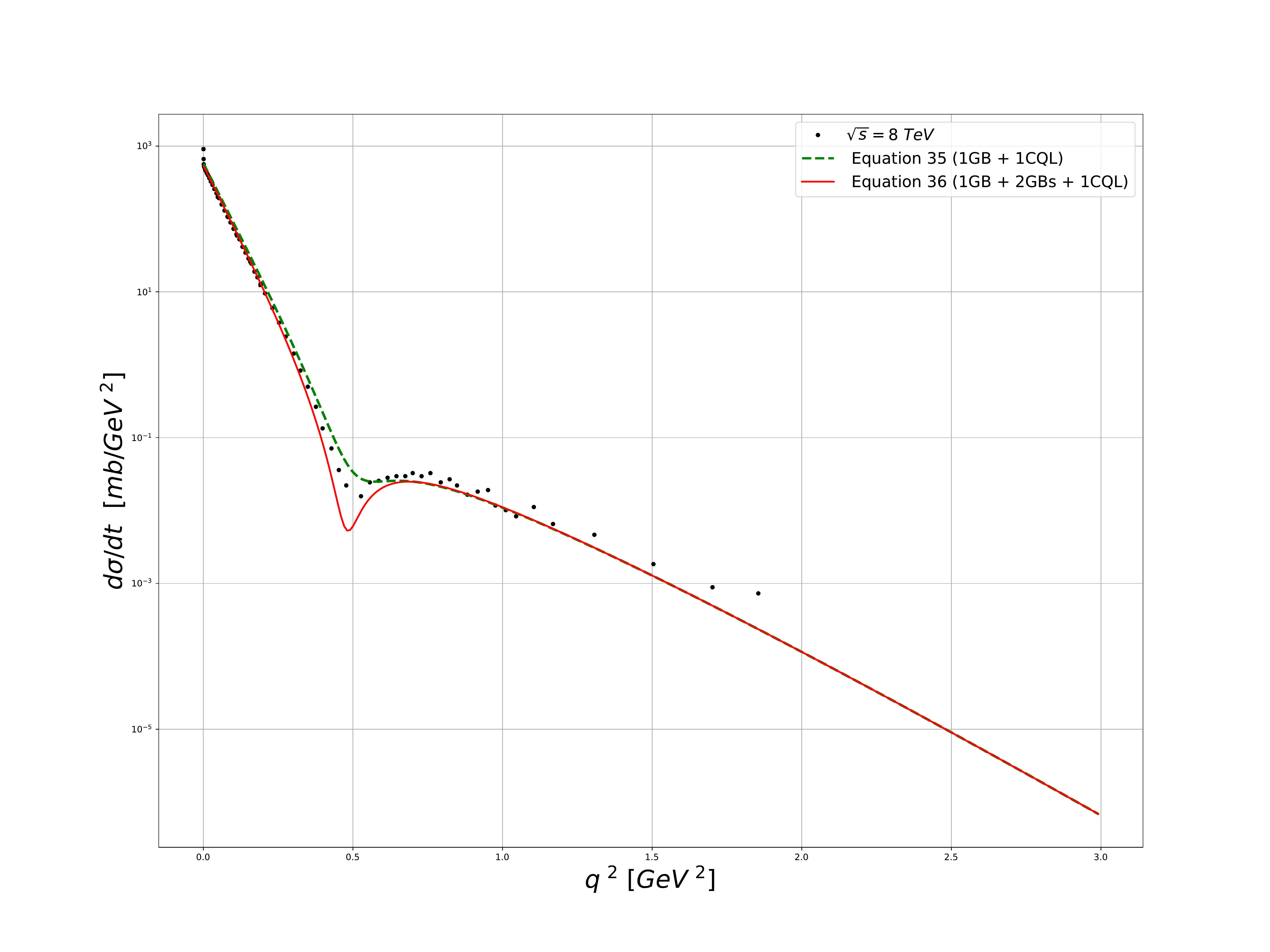}
\caption{Elastic pp scattering differential cross section at $\sqrt{s} = 8$ TeV. Black dots are experimental data, dashed line is the result of eq.(35), solid line comes from eq.(36). }
\label{8tev}       
\end{figure}

\begin{figure}
\centering
\includegraphics[width=12cm]{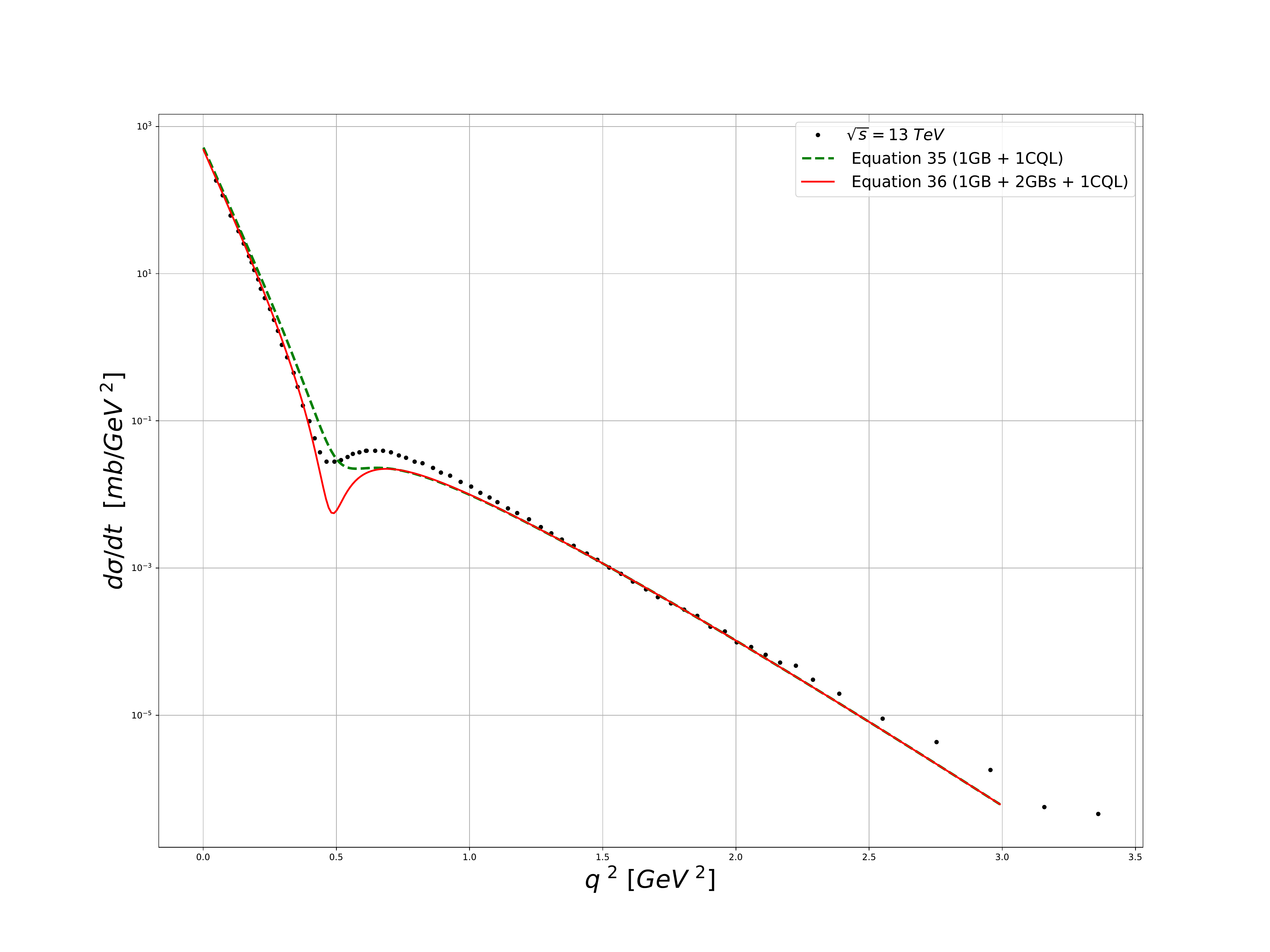}
\caption{Elastic pp scattering differential cross section at $\sqrt{s} = 13$ TeV. Black dots are experimental data, dashed line is the result of eq.(35), solid line comes from eq.(36). }
\label{13tev}       
\end{figure}

\bigskip

2) The total cross section.

\medskip

From eq.(33) and/or eq.(34), we obtain for the total cross section for $pp$ scattering:
\begin{equation}
\sigma_{tot}(s) = \frac{4M^{2}}{s}\,{\rm Im} \,T(s, \vec q = 0) = \frac{g}{\sqrt2}\,\delta_q^2 = \frac{g}{\sqrt2}\,\Bigl({\lambda\over m}\Bigr)^2\,(6m)^{2p}\,s^{-p}
\end{equation}

This cross section comes from the exchange of one gluon bundle only, between the two protons, the contribution of the chain to Im\,$T$ giving 0 at $q^2 = 0$. Of course, we have taken the smallest possible number of bundles and chains to do our computations. Higher terms in the expansion of eq.(30) would modify $\sigma_{tot}$,  and also we need to know more precisely how all the parameters of our model vary with $s$ to give a more reliable prediction on how $\sigma_{tot}$ varies with energy.

\bigskip

{\section{Summary and expectations}}

We have given a theoretical description of the $pp$ scattering differential cross sections, using the simplest two quantities our model can give: one and/or one plus two gluon bundles,  plus one chain with one loop, to be exchanged between two quarks, done in the eikonal frame, and with a lot of simplifications in our formulas. The forms and results of the above calculations and data fits can be improved by several manners. 
\medskip

1) We can expand the quark loop chain term ( eq.(25) ) to higher orders, and we can also consider chains with more than one loop ( see, for instance Ref.\cite{qcd4}, eq.(31) ). In fact, the small loop renormalization parameter $\kappa$ defines a possible perturbative sequence: it appears to be so small that it might be used to systemically neglect higher numbers of closed quark loops and closed quark loops chains and gluon bundles interferences. Whatever, in this scattering problem, one must retain at least one loop, in order to show a minimum followed by a $q^2$ dependent rise and then fall of the differential cross section with increasing $q^2$. We can also add more gluon bundle exchanges to the amplitudes ( eq.(24) ). As can be seen on the figures, taking into account a two bundle exchange deepens the dip and give a better shape for fitting the data. We have chosen to plot both curves on the same figure with the same averaged parameters, so one can see the differences produced between one and two bundles. And as can also be seen from these parameters, we need larger $\lambda$ and $\kappa$ for the LHC, which can mean that we need more bundles and loops to produce better fits to the TOTEM data, and get $\lambda$ and $\kappa$ comparable to those of the ISR data. We also have to take into account the wide range of energy for the TOTEM results.
\medskip

2) We have to work on the energy dependence of our model which appears clearly when we go from ISR to LHC data. In particular, the major ingredient of our formulas, $\varphi(\vec b)$, should probably be changed to a $\varphi(\vec b, E)$, to justifie the variation with energy of our masses $m$ and $\bar m$. Of ourse, all our parameters have an energy dependence which has to be understood. Appendix A and B give insight of how to treat the $f\!\cdot\!\chi$  term in its globality, instead of using it's magnitude $R$, and recover some energy dependence.
\medskip

3) Finally, the problem of replacing a six--body problem by a two--body one, as we do in this article, can be be improved by random matrix technics, as mentioned in Appendix B.
\medskip

To end this article, a comment on the suggested appearance of Pomerons, resulting from our non--perturbative analysis may be appropriate. An immediate statement is that, in no way, are our results specifically related to any of the many perturbative calculations and Reggeon estimations of soft and hard Pomerons \cite{DHM,DL}; but we do find a natural separation of our amplitudes and differential cross sections into a dominant part at small momentum transfers, and another part which becomes important at larger momentum transfers. For example, for $q^2$ values less than the dip position, the contribution of one or two gluon bundle terms is dominant; while rising from zero, and for $q^2$ values larger than that of the dip, it is the closed loop chain which plays the dominant role. If one wishes to use Pomeron terminology, one can refer to these respective contributions as "non-perturbative soft and hard Pomerons".

\bigskip

\bigskip
{\bf\section* {Appendix}}
\def\theequation{{A}.\arabic{equation}}
\setcounter{equation}{0}
\appendix

{\section{The substitution of $(f\!\cdot\!\chi)_{\mu\nu}^{ab}$ by $R$}}
\setcounter{equation}{0}

In this Section it is argued that the simplification taking the matricial structure $f\!\cdot\!\chi$ to the simple real scalar $R$ of eq.(10) and $R_1$ and $R_2$ of eq.(15), however drastic, is able to preserve the essential dynamical content of eq.(8).
\par\smallskip
As an example, we start from the expression of $I(g^2,q^2)$ of eq.(31) in Ref.\cite{qcd4}:
\begin{equation}
I(g^2,q^2)=\mathcal{N}\int \mathrm{d}\chi(O)\,\det(gf\cdot\chi(O))^{-\frac{1}{2}}\,e^{\frac{i}{4}\chi^2(O)}\, \frac{(f\cdot\chi(O))^2}{(f\cdot\chi(O))^2+(\lambda\kappa g q^2\tilde{\varphi}(q))^2}\end{equation}

A random matrix treatment of the functional integration on $(f\!\cdot\!\chi)^{ab}_{\mu\nu}(O)$ can be performed along the lines of \cite{qcd5, EL1, EL2}, with, as a result ($X=\lambda\kappa g q^2\tilde{\varphi}(q)$):
\begin{equation}
I(g^2,q^2)=-\mathcal{N}\,X^2\,\biggl\langle \pm\sum_{\{q_i\}}\prod_{i=1}^N\int \frac{\mathrm{d}\xi_i}{\sqrt{\xi_i}}\,\xi_i^{q_i}\,e^{\frac{i}{8N_c}\xi_i^2}\, \mathrm{diag}\,\left(\dots, \,\frac{1}{\xi_j^2+X^2},\,\dots\right)\biggr\rangle_{O_N(I\!\!R)}\end{equation}
where the sum ranges over the monomials of a Vandermonde determinant $\prod_{1\leq i<j\leq N} (\xi_i-\xi_j)$, the $q_i$-powers satisfying the constraint of an equal global degree of $\sum_{i=1}^N q_i=N(N-1)/2$.
 
\medskip
\noindent In (A.2), the large brackets stand for an average value prescription taken over the orthogonal group $O_N(I\!\!R)$:

\begin{equation}
{\biggl\langle A\biggr\rangle}_{O_N(I\!\!R)} = {\mathcal{N'}}^{-1}\int \mathrm{d}\mathcal{O}\ {}^t\mathcal{O}\,A\,\mathcal{O}\,,\ \ \ \ \mathcal{O}\in O_N(I\!\!R)\end{equation} 
where $\mathcal{N'}$ is the orthogonal group volume. As demonstrated in \cite{EL1, EL2}, the `color angular' degrees of freedom then decouple from the integrations over eigenvalues (the $\xi_is$), and in the general case, factor out the $SU(3)$ color Casimir invariants dependences.
\par\medskip
The dynamical aspect which depends on the coupling constant $g$, that is on $X$ (which factorizes (A.2) at the squared power), comes about with the integrations:
\begin{equation}
\int_\infty^{+\infty}\frac{\mathrm{d}\xi_j}{\sqrt{\xi_j}}\,\xi_j^{q_j}\,e^{\frac{i}{8N_c}\xi_j^2}\,\frac{1}{\xi_j^2+X^2}\end{equation}

Since, moreover, $|X|<<1$, one can see how eq.(8) is connected to the basic dynamical piece (A.2) of the exact integration process, and is therefore able to capture the qualitative features of (A.2) and (A.4) in the much simplified way followed in the current paper.

\bigskip

\def\theequation{{B}.\arabic{equation}}
\setcounter{equation}{0}
{\section{Basis of $E$-dependences}}

We see that our parameters have a slight dependence on energy, visible as it increases from ISR to LHC values; and that such changes could be due to our two--body description of this six--quark scattering reaction, and, also, to the degree of arbitrariness of $\varphi(\vec b)$.
\par
The above simplification, together with the `shrinkage' of the basic structure of $f\!\cdot\!\chi$ into the only real variable $R$ doesn't allow one to keep track of the scattering energy dependences. This is why, as stated above, the functions $(m/E)^{2p}$ of (33) and (34) can only be dictated by a fit to experimental data.
\par
However, it is worth noticing that the random matrix exact treatment of the basic $f\!\cdot\!\chi$ structure sheds some light on this issue. In strong coupling, eikonal and quenching approximations at least, the two by two scattering process is controlled by an expression ( ref\cite{EL2}, eq(18) ):

\begin{eqnarray}\label{new1}
&& \nonumber\pm\!\! \sum_{\mathrm{monomials}}\biggl\langle\,\prod_{i=1}^N\, \biggl[{{\sqrt{4iN_c}}\,{\sqrt{{\widehat{s}}({\widehat{s}}-4m^2)}}\over {m^2}}\biggr]\, \frac{[({\cal{OT}})_i]^{-2}}{g\varphi(b)}\\ &&\times\, G^{30}_{03}\!\left( \biggl[{  g\varphi(b)\over {\sqrt{32iN_c}}  }{m^2\over {\sqrt{{\widehat{s}}({\widehat{s}}-4m^2)}}}\biggr]^2\biggl[({\cal{OT}})_i\biggr]^{4}\, \biggr|\frac{1}{2}, \frac{3+2q_i}{4},1\!\right)\,\biggr\rangle_{\!O_N(\mathbb{R})}
\end{eqnarray}
where the sum runs over the monomials of a Van der Monde determinant expansion, each of them characterised by a distribution of powers $\{\dots,q_i,\dots\}$ such that $\sum_1^N q_i=N(N-1)/2$. Eq.(B.1) accounts for the energy dependence of a two body scattering process (in here, two quarks of the same flavour). Though expressed in terms of kinematical invariants, $m^2$ and $\hat{s}=(p_1+p_2)^2$, the $G^{30}_{03}$--Meijer function argument is written in the center of mass system of the colliding quarks of momenta $p_1$ and $p_2$. In (B.1), $N=D\times(N_c^2-1)=32$, is the full format of the matrix representation of the structure $(f\!\cdot\!\chi)$. As the average on the orthogonal group is taken, the Casimir invariant dependences factor out, the same for each monomial of the sum (B.1). At leading order, one gets:
\begin{equation}
\pm \frac{DC_{2f}}{N} I\!\!\!I_{3\times 3}\sum_{\mathrm{monomials}}\ (\prod_1^NA^1_i)\ (\sum_1^N\frac{A^3_i}{A^1_i})\biggl\lbrace {g\varphi(b)\over {\sqrt{2iN_c}}}\times (\frac{\mathbf{m}^2}{{\hat{\mathbf{E}}}^2})\,, \ \ \ \ g\frac{\mu m}{|u'_0\bar{u}'_3|}\, e^{-(\mu b)^2}\ \times(\frac{\mathbf{m}}{\hat{\mathbf{E}}})         \biggr\rbrace \end{equation}
the two relations of $m/\hat{E}$ depending on how one defines and uses the eikonal approximation. The mass term $\mu$ is the mass scale associated to the property of effective locality. The coefficients appearing in (B.2) are well defined by the analytic properties of the Meijer special functions \cite{EL2}:

\begin{equation}\label{Anumbers}
A^1_i=\Gamma(\frac{1}{2})\Gamma(\frac{2q_i+1}{4}),\  \ \ \ A^2_i=\Gamma(\frac{1-2q_i}{4})\Gamma(\frac{-2q_i-1}{4}),\  \ \ \ A^3_i=\Gamma(-\frac{1}{2})\Gamma(\frac{2q_i-1}{4})
\end{equation}

At next to leading orders, extra $C_{3f}$ Casimir operator dependences show up, as well as other energy dependences of forms $(m/\hat{E})^n$ with $n\geq 2$. Important remarks are in order.
\par
\medskip
\begin{itemize}
\item Inspection shows that the monomials of (B.1) appear with an equal number of $+$ and $-$ signs, and that the particular law of $(m/E)^{2p}$, which in (33) and (34) comes about to the squared power, and cannot be derived as such, out of (B.1), (B.2) or any further expansions thereof. 
 \item
 Now, alternatively, one may look at the theoretical value of $p$, identified out of (B.2) at $p=0.13$ and $p=0.055$, as not so bad a result in view of the approximations and simplifications adopted; and in particular, a value able to account for qualitative features of the ISR curves which couldn't be explained otherwise.
 \item
 The energy dependence of  $(m/E)^{2p}$ extracted from the data, can be supported at the theoretical level if they represent a numerical fit to the involved expansions (B.2). The latter however comprise such a large number of monomials (as much as $2^{120}$ terms in some symmetric situations and $2^{496}$ terms otherwise!), that no computer could possibly probe such a `fitting test'. This should be examined on the simpler available situation.
 \end{itemize}
 \par
\medskip
For example, the unphysical case of $N=4$ comprises already $2^6=64$ terms, to wit \cite{EL2}:

\begin{eqnarray}\label{monomials'}&&+(3210)-(3201)+(3102)-(3120)+(3021)-(3012)-(2211)+(2202)  \nonumber \\ &&-(2103)+(2121)-(2022)+(2013)-(2220)+(2202)-(2112)+(2130)\nonumber \\ &&  -(2031)+(2022)+(1221)-(1212)+(1113)-(1131)+(1032)-(1023) \nonumber \\ &&-(2310)+(2301)-(2202)+(2220)-(2121)+(2112)+(1311)-(1302)\nonumber \\ && +(1203)-(1221)+(1122)-(1113)+(1230)-(1221)+(1122)-(1140)\nonumber \\ &&     +(1041)-(1032)-(0231)+(0222)-(0123)+(0141)-(0042)+(0033)
\end{eqnarray}
where the monomials are represented by the powers $q_i$ written seqentially: $+(2301)$ stands for the result of the integration over $\xi_1,\xi_2,\xi_3$ and $\xi_4$ of the monomial $+\xi_1^{2}\xi_2^{3}\xi_3^{0}\xi_4^{1}$. Obvious cancellations leave only 6 residual monomials \cite{EL2}:
\begin{equation}\label{final'}2\times(0222)-2\times(0123)-(1212)+(0141)-(0042)+(0033)\end{equation}

An encouraging indication would be to see whether the six residual monomials of (B.5) produce an overall multiplicative factor such that, multiplying a behaviour of $(m/E)$, the net result is eventually mimicked numerically by a depleted behaviour of $(m/E)^k$ with $k<1$, at least over some (ISR) range of energy values. This amounts to the following transcription:
\begin{equation}
\frac{DC_{2f}}{N} I\!\!\!I_{3\times 3}\ \biggl[\frac{g\mu m}{|u'_0\bar{u}'_3|}\, \ \sum_{res.mon.}(\prod_1^4A^1_i)\ (\sum_1^4\frac{A^3_i}{A^1_i})\biggr]\ (\frac{m}{E})\longrightarrow \frac{DC_{2f}}{N} I\!\!\!I_{3\times 3}\ (\frac{m}{E})^k
\end{equation}
where the sum runs over the residual monomials of (B.5). A necessary condition for this numerical fit to be relevant is that the bracket of (B.6) be larger than one, strictly. Keeping the same value of $g$, and the values $m=5$ MeV, $\mu\simeq \sqrt{\hat{s}}=120$ MeV (that is ${|u'_0\bar{u}'_3|}\simeq \hat{s}$, which is the eikonal evaluation), one finds first:
\begin{equation}
\sum_{res.mon.}(\prod_1^4A^1_i)\ (\sum_1^4\frac{A^3_i}{A^1_i})=16,743\  \pi^2\end{equation}
and eventually:
\begin{equation}
\biggl[\frac{g\mu m}{|u'_0\bar{u}'_3|}\, \ \sum_{res.mon.}(\prod_1^4A^1_i)\ (\sum_1^4\frac{A^3_i}{A^1_i})\biggr]\simeq 6\,\pi^2
\end{equation}

Even though going in the right sense, this large number would select a (too) small value of $k$, over a (too) restricted range of (ISR and LHC) energy values. What must be kept in mind, though, is that (B.7) holds at the \textit{partonic} level, on the one hand (see below), and that on the other hand, the value of $N=4$ which is here taken as the simpler tractable example is certainly very far from the physical case of $N=32$.
\par\medskip
This however sheds on the parameter $p$ the following light. Even at eikonal, quenching and large coupling limits, a tight enough control of energy dependences is guaranteed by Eq.(B.1) which enjoys a rigorous derivation \cite{qcd5}. The matter though, at the physical value of $N=D\times(N_c^2-1)=32$ is the number of monomials generated by a Van der Monde determinant : As much as $2^{120}$ terms at least, each of them contributing at orders $(m/E)$, $(m/E)^2$, etc... As suggested by the simplest example of $N=4$, such a sum can be fitted numerically by an overall dependence of form $(m/E)^{2p}$, at least over some range of (ISR and LHC) energy values. No computer could otherwise deal with so many terms.
\par
Incidentally, it can be checked also that passing from $\hat{s}$ to $s$ variables, \textit{i.e.} from quarks to protons doesn't bring drastic changes to the conclusions above so long as the ratio $M_p^2/E^2$ is much smaller than 1, where $M_p^2$ is the squared proton's mass ($\simeq 1GeV^2$), a condition amply satisfied in the range of ISR and LHC energies (it simply seems to diminish the value of the bracketed quantity of (B.8)). 

In eq.(83--84) of Ref.\cite{qcd5}, a single quark--quark scattering subprocess produced an energy dependence of $(m/\sqrt{\hat s})^{2}$. This result is in line with the phenomenologically introduced energy dependence factor of $(m/E)^{2p}$ in eq.(33) and (34) above; but it is no way its justification. A proper derivation of the correct energy dependence for elastic, high energy $pp$ scattering awaits the more complicated 6--quark scattering and rebinding analysis, in which six sequences of Meijer G--functions are combined into two final protons, a six--body problem which we have avoided by our use of a two--body approximation, requiring our phenomenological energy dependence. 

\bigskip
\bigskip

\noindent{\bf\Large Acknowledgement}
\vskip0.3truecm 
This publication was made possible through the support of a Grant from the Julian Schwinger Foundation. We especially wish to thank Mario Gattobigio for his many kind and informative conversations relevant to the Nuclear Physics aspects of our work. It is also a pleasure to thank Mark Rostollan, of the American University of Paris, for his kind assistance in arranging sites for our collaborative research when in Paris.

\end{document}